# Integrated Platform for Robust Differential Refractive Index Sensor

Andre L. Moras,[1,*] Valnir C. S. Junior,[2] Mario C. M. M. Souza,[1] Giuseppe A. Cirino, [2] Antonio A. G. Von Zuben,[1] Newton C. Frateschi,[1] Luis A. M. Barea,[1,2]

[1] *Gleb Wataghin Physics Institute, University of Campinas, Campinas, SP Brazil*
[2] *Dept. of Electrical Engineering, Federal University of São Carlos, São Carlos, SP, Brazil*
*Corresponding author: morasal@ifi.unicamp.br

**In this work, we demonstrate an integrated platform comprising a refractive index (RI) sensor based on Photonic Molecule (PM) that effectively mitigates the influence of environmental perturbations using a differential measurement scheme while providing high quality-factor (Q) resonances. The RI sensor consists of an exposed microdisk resonator coupled to an external clad microring resonator fabricated on silicon-on-insulator (SOI) platform. We report a RI sensitivity of 23 nm/RIU, achieving a limit of detection (LOD) of $1.6 \times 10^{-3}$ refractive index units (RIU) with improved stability in a compact footprint of $40 \times 40$ μm², representing a good solution for real-life applications in which measurement conditions are not easily controllable.**

In the past decades, there have been enormous advances in the research of SOI-based integrated optical microcavities fueled by the high scientific interest and technological demands in several areas including optical communications and optical sensing [1–4]. These optical microcavities such as rings or disks allow strong light confinement and their dimensions can be adjusted for a given desired spectrum and application [2,5]. When these microcavities are made of silicon, the tight light confinement within the cavities allows them to be ultra-compact and easy to integrate with other devices [2,4]. Harnessing these advantages, the cavities can compose very large scale integrated optoelectronic circuits, fabricated by mature microfabrication technology based on complementary metal-oxide-semiconductor (CMOS) [6]. In particular, these cavities also ensure high sensitivity to refractive index (RI) changes and potential integration with micro-fluidic applications [3,7–10]. For this, they allow the implementation of gas, temperature and biological material sensing functionalities on one chip, the so-called lab-on-a-chip (LOC) [8,11,12], permitting significant advances in fundamental areas as environmental monitoring and biomedicine [13–17].

Several studies in the literature show the efficiency of SOI microcavities coupled to waveguides to monitor dynamic molecular reactions, quantitative concentrations of solutions and to determine chemical affinities for medical studies and clinical diagnosis [4,8,18]. These studies usually show that a microcavity can be used as optical sensor in homogeneous sensing approach when the analytes, suspended in a fluid medium, are covering the microcavity top. In this situation, the effective index ($n_{eff}$) of the cavity mode will change, resulting in a corresponding deviation of the resonance wavelength, guaranteeing the sensing capacity [2,7,11].

In the literature, a widely accepted figure of merit used to calculate the performance of optical sensors is the resonator RI sensitivity ($S_{RI}$) that can be defined as [7,19,20]:

$$S_{RI} = \frac{\Delta \lambda_{res}}{\Delta n_c} = \left( \frac{\lambda_{res}}{n_g} \right) S_w \qquad (1)$$

where $\Delta n_c$ is the change in the cladding refractive index over the cavity, $\Delta \lambda_{res}$ is the change in the resonant wavelength due to $\Delta n_c$, $\lambda_{res}$ is the resonance wavelength, $n_g$ is the group index and $S_w$ is the waveguide sensitivity, defined as:

$$S_w = \frac{\Delta n_{eff}}{\Delta n_c} \qquad (2)$$

The resonator RI sensitivity is given as the product of a waveguide intrinsic characteristic ($S_w$) and a design dependent feature ($\lambda_{res}/n_g$) introduced by the resonator. Another important figure of merit is the limit of detection (LOD) [7] measured in refractive index units (RIU):

$$LOD = \frac{\lambda_{res}}{Q \, S_{RI}} \qquad (3)$$

The LOD indicates the minimum detectable concentration changes. A small LOD is desired so that small variations in $n_c$ can be detected by a deviation in $\Delta \lambda_{res}$. This requirement can be satisfied by maximizing the Q or the $S_{RI}$ of the resonator. In fact, when $n_c$ is changed, $n_{eff}$ also undergoes a change, so that for a small wavelength range, $\Delta n_{eff}$ is practically proportional to $\Delta n_c$ as:

$$\Delta n_{eff} = K \cdot \Delta n_c \qquad (4)$$

where the parameter K is constant and depends on the material and waveguide cross-section [7,20]. The relationship between $\Delta \lambda_{res}$ and $\Delta n_c$ can then be written as:

$$\Delta n_c = \frac{n_g}{K} \frac{\Delta \lambda_{res}}{\lambda_{res}} \qquad (5)$$

A big problem regarding sensors based on optical resonators with high-Q is that they are also highly sensitive to fluctuations induced by the environment, such as temperature and air humidity variations, which compromises the reliability and stability of these sensors. One solution to overcome these limitations and improve the sensor stability and reliability is presented in this work, where it is suggested to build optical sensors based on photonic molecules (PMs) [21,22]. In particular, we present here the design, fabrication and characterization of a sensor based on a PM, which consists of an exposed microdisk resonator coupled to an external clad microring resonator.

In the literature, it has been shown that when multiple resonators are coupled, it is possible to form compact PMs [23,24,24–26]. The complexity of PM's transmission spectrum will depend on the number of coupled resonators and how they couple to each other. The transmission spectrum of a single microring resonator in an add-drop filter (Fig. 1(a)) presents resonances separated by the free spectral range (FSR), which is inversely proportional to the resonator length, and characterized by the linewidth (FWHM) and the extinction ratio (ER). When a second distinct resonator, such as a disk, is coupled to the first one (Fig. 1(b)), the two resonators may have different resonance wavelengths (non-degenerate condition) or they can be both resonant at the same wavelength (degenerate condition). In the non-degenerate condition, the resonances of the outer ring remain unaffected and a new resonance appears which is associated with the embedded disk. When the two resonators are degenerated, however, their mutual coupling induces mode-splitting proportional to the coupling strength between resonators.

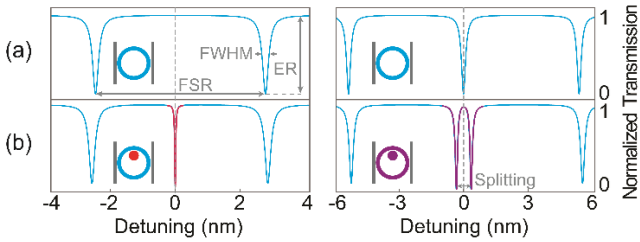

**Fig. 1.** Comparison between the transmission spectrum of a single microring resonator (a) and a PM based on two resonators (b) for both non-degenerate (left side) and degenerate (right side) conditions. The transmission spectra are shown as a function of the detuning with respect to the embedded disk resonances [25]. The colored resonances are associated with each one of the cavities in the inserted schemes.

Fig. 1(b) shows that in the non-degenerate condition, the notch resonance (red resonance) associated with the embedded disk has a Q higher than the Q associated with the external-ring resonance. This fact occurs due to the reduction in the coupling losses between the inner disk and the straight waveguide in this kind of PM, in which the waveguide-disk coupling is always intermediated by the external ring. For this reason, the non-degenerate situation allows the use of the disk resonance (called detection resonance, $\lambda_{detection}$) for accurate detection and its wavelength shifts ($\Delta\lambda`_{before}$ and $\Delta\lambda`_{after}$) may be measured in relation to the external ring resonances (called reference resonances, $\lambda_{ref1}$ and $\lambda_{ref2}$). These reference resonances should always be present and need to be fixed and free of deviations caused by the cladding refractive index changes, allowing the differential measurement between the detection resonance and these reference resonances ($\Delta\lambda_{res} = \Delta\lambda`_{after} - \Delta\lambda`_{before}$) in a compact footprint and with improved stability.

However, the measurement range of these sensors is dependent of the wavelength shift $\Delta\lambda_{res}$, which is limited by the free spectral range (FSR) of the external ring resonator, such that

$$\Delta\lambda_{res} < FSR = \left(\frac{\lambda_{res}^2}{n_g L}\right) \qquad (6)$$

where L, in the proposed device, is the length of the outer ring resonator (L = $2\pi$R, where R is the resonator radius). Therefore, comparing to eq. 5, the measurement range of $n_c$ is then given by

$$\Delta n_c < \frac{\lambda_{res}}{K L} \qquad (7)$$

Fig. 2(a) depicts a schematic view of the planned optical sensor based on PM that meets the requirements above. The sensing window is open on the top of the disk, exposing only half of its upper surface, and allows the detection resonance to suffer deviations (Fig. 2(b)) associated with changes in the cladding refractive index. These deviations can be caused by a fluid (either gas, air or water), which has been loaded on the window.

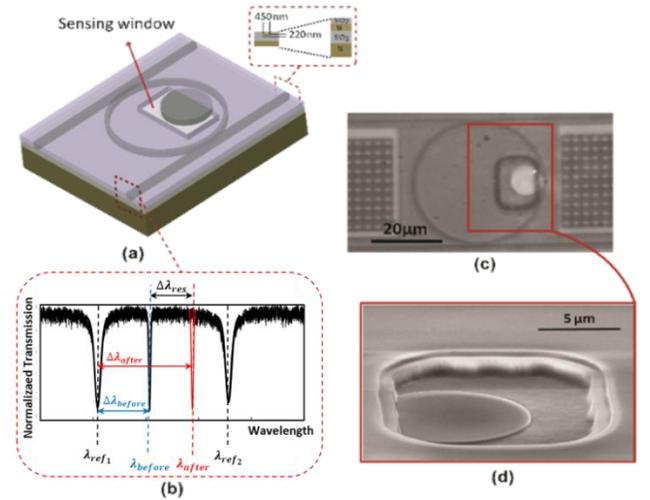

**Fig. 2.** (a) Schematic view of the sensor based on PM. (b) Desired spectrum for this sensor. The red curve shows a deviation of the resonance of the disk caused by a change in the cladding refractive index. (c) Optical micrograph of the fabricated sensor and (d) Scanning electron micrograph of the sensing window.

This sensor was fabricated by using a SOI platform at IMEC-EUROPRACTICE. The ring and disk radius are 20 µm and 5 µm, respectively. The gaps between waveguide-ring and ring-disk are 200 nm. The effectively single mode (TE) waveguides consist of a silicon core with 220 nm x 450 nm cross-section on a 2-µm-thick layer of buried thermal SiO2, and it is covered with a 1-µm thick deposited SiO2 layer. The sensing window over part of disk was defined by photolithography and etching process (Fig. 2(c) and Fig. 2(d)) and the chips containing the sensors were spliced in areas where there were inverse nanotapers for efficient coupling to optical fibers. Polishing technique with focused ion beam (FIB) milling was employed to obtain high-quality mirrors with excellent coupling condition [27,28].

The spectral features of the sensor were carried out using 2-dimensional finite-difference time-domain (2D-FDTD) algorithm of the FullWAVE package. The 3D structures of the sensor were reduced to 2D structures, employing the effective index method [20] and extensive numerical simulations with the 3D beam propagation method (BPM) included in the BeamPROP package. It is mandatory to compute this reduction to decrease the simulation time and the required computing memory. The spectral response performed in the through port of the sensor is shown in Fig. 3(a). A schematic representation of the simulated sensor is inserted in this figure, showing that only half of the inner disk surface can sense changes in the cladding refractive index (blue part of the disk). The red parts in this scheme represent the waveguides, ring and half of the inner disk. These parts are buried in a SiO₂ cladding.

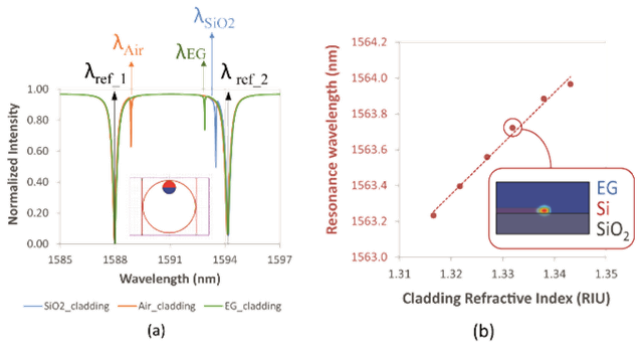

**Fig. 3.** (a) Spectral features of the sensor for three conditions of cladding materials over the half-disk: SiO₂ (blue line), Ethylene Glycol (green line) and air (orange line). Inset: schematic view of the sensor highlighting the part of the disk responsible for the sensing (blue part). (b) Resonance wavelength simulated at disk for a range of cladding refractive index. The figure inset shows the calculated spatial distribution of electrical field radial component in the cross-section of the sensing disk.

To perform the simulation, three conditions were considered for $n_c$ on the half-disk (blue part): SiO₂ (blue curve), air (orange curve) and Pure Ethylene Glycol (EG) (green curve). The spectral responses of the sensor shows that the resonances associated with the ring are always fixed in the same wavelengths ($\lambda_{ref1}$=1587.99 nm and $\lambda_{ref2}$=1594.15 nm), regardless of the cladding material used over the half-disk. The detection resonances suffers a redshift when the $n_c$ is increased, for example, from air ($n_c$=1.000) to SiO₂ ($n_c$=1.444) [21], both for $\lambda$=1.59 µm. These resonances are $\lambda_{air}$=1588.86nm, $\lambda_{EG}$=1592.90 nm and $\lambda_{SiO2}$=1593.50 nm. The deviations between these resonances and one of the reference resonances ($\lambda_{ref1}$) can be computed as $\Delta\lambda' = |\lambda_{detection} - \lambda_{ref1}|$, where $\lambda_{detection}$ is each one of the detection resonances. The differences for each case are $\Delta\lambda'_{air} = 0.87$ nm, $\Delta\lambda'_{EG} = 4.91$ nm and $\Delta\lambda'_{SiO2} = 5.51$ nm.

The sensitivity of this sensor can be computed according to Eq. 1 and the change of the resonance wavelength $\Delta\lambda_{res}$ can be found by applying the relation $\Delta\lambda_{res} = |\Delta\lambda'_{SiO2} - \Delta\lambda'_{EG}|$ for the case where the sensitivity is associated with the change of $n_c$ from $n_c^{EG} = 1.420$ to $n_c^{SiO2} = 1.444$. The advantage of computing $\Delta\lambda_{res}$ in this way is that $\Delta\lambda'$ will always be present in the spectra, regardless of deviations caused by imprecision during the data acquisition. The calculated change of the resonance wavelength is $\Delta\lambda_{res} = 0.6$ nm and the sensor RI sensitivity is $S_{RI}$~25 nm/RIU. Fig. 3(b) shows the simulated resonance wavelength against the refractive index of the cladding for a fully exposed disk. In this case, the linear fitting shows that the RI sensitivity is increased to ~28 nm/RIU. The inserted figure shows the calculated spatial distribution of the electrical field radial component in the cross-section of the sensing disk.

The experimental characterization of the fabricated sensor was performed as follows. A tunable laser (1465 nm to 1640 nm) was used as light source and the coupling of light into the waveguide was done with GRIN rod lensed optical fibers. The transmitted light was collected using similar lenses and sent to an InGaAs power meter. The temperature was stabilized at 25 °C and the input polarization was controlled to allow efficient coupling to the quasi-TE mode of the silicon waveguide. Fig. 4(a) shows the experimental transmission spectra of the sensor obtained for pure water and five aqueous EG solutions with different volume concentrations. In each case, the transmission spectrum of the sensor was measured after a droplet of approximately 0.5 µl of the prepared solution was applied on the sensor surface. The same experiment was repeated five times for each solution to secure the reliability of the developed sensor. The refractive indices of the aqueous EG solutions were obtained from [29].

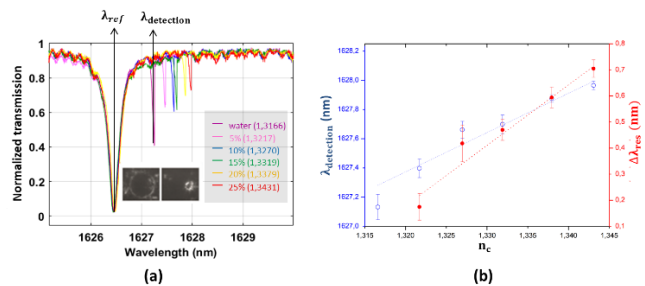

**Fig. 4.** (a) Experimental spectra of the sensor for pure water and five different aqueous EG solutions. The inset figures are the infrared microscopy images of the scattered light in the resonances $\lambda_{ref}$ and $\lambda_{detection}$ respectively; (b) Resonance wavelength positions (blue circles) and sensor response (red dots) as a function of the cladding refractive index. The dashed lines are linear fittings.

The spectral responses of the sensor (Fig. 4(a)) shows that the ring reference resonance remained fixed ($\lambda_{ref} \cong 1626.5$ nm) during the experiments. The inset figures show the corresponding infrared microscopy images of the scattered light in the resonances $\lambda_{ref}$ and $\lambda_{detection}$ (approximately 1628 nm). These images demonstrate that when the laser is pumping in each resonant wavelength, the spatial distribution of the optical mode will be concentrated in the cavity associated with the resonance, as expected for a PM operating in the non-degenerated condition. During the sensor operation, the detection resonance suffered a redshift when the concentration of EG in water was increased. This redshift was caused by the increase of the $n_c$ from 5% ($n_c$=1.3217) to 25% ($n_c$=1.3431) and it is shown in details in Fig. 4(b) (blue circles).

The deviations between these detection resonances and $\lambda_{ref}$ were calculated from $\Delta\lambda' = |\lambda_{detection} - \lambda_{ref}|$ and the sensor response ($\Delta\lambda_{res} = \Delta\lambda'_{after} - \Delta\lambda'_{before}$) is also presented in Fig. 4(b) (red curve). The slope (RI sensitivity) of the linear fitting shows an experimental sensitivity of $S_{RI} = (23.2 \pm 2.2)$ nm/RIU, giving a limit of detection LOD ~ 1.6×10⁻³ RIU. This result is in good agreement with our theoretical prediction for the sensitivity and is below the $S_{RI} = 27.1$ nm/RIU presented in reference [7] to a single disk sensor (R = 5 µm). However, our PM-based sensor has only half of the disk susceptible to a change of $n_c$ and it allows the presence of reference resonances requiring a reduced footprint (40 x 40 µm²).

The sensor response to temperature variation was also evaluated by employing a temperature controller setup composed of a thermostat (ILX-Lightwave – LDT-5525– Laser Diode Temperature Controller) and a Peltier (CP1.0-63-08L from Melcor). Fig. 5(a) shows the experimental transmission spectra obtained for four different temperatures. As expected, both the reference and detection resonances suffered a redshift when the temperature was increased. However, Fig. 5(b) shows that the resonance spectral shifts for the reference ring (blue curve) and detection disk (red curve) presents different sensitivities to temperature variations, $(0.081 \pm 0.001)$ nm/°C and $(0.087 \pm 0.003)$ nm/°C, respectively. Their sensitivities to temperature variations are not equal because the reference and detection cavities have different radius and also because only half part of the disk was exposed to water/solution cladding. Now, considering the spectral shift of the sensor response ($\Delta\lambda_{res}$, black curve), its sensitivity to temperature variation is reduced to $(0.008 \pm 0.002)$ nm/°C. This result shows that this PM-based sensor is about 10 times less sensitive to temperature variations than its reference and detection absolute resonances or others sensors based on single cavities.

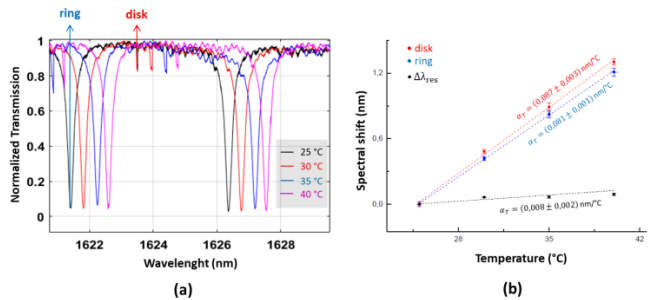

**Fig. 5.** (a) Experimental spectra of the sensor measured for four different temperatures; (b) Spectral shifts for the reference ring (blue curve) and detection disk (red curve) resonances as function of chip temperature. The black line represents the spectral shift of the sensor response ($\Delta\lambda_{res}$) against chip temperature. The dashed lines are linear fittings with its related slopes.

In conclusion, we have demonstrated here the design, fabrication and characterization of a sensor based on a PM, that is composed of a microdisk internally coupled to a microring in a well-known add-drop filter configuration. In this sensor, only the inner microdisk is sensitive to changes in the top cladding refractive index and its resonance, non-degenerate with the microring resonances, has a high-Q (~40,000). One of these non-degenerate resonances was employed in the detection of different aqueous ethylene glycol solutions by measuring its deviation from the reference resonances of the microring. An experimental RI sensitivity of $(23.2 \pm 2.2)$ nm/RIU was achieved with a LOD of approximately $1.6 \times 10^{-3}$ RIU in a compact footprint (40 μm x 40 μm). Also, it was shown that the sensor response to temperature variations is about 10 times lower than the individual responses of the reference ring and the detection disk.

**Funding.** This study was financed in part by the Coordenação de Aperfeiçoamento de Pessoal de Nível Superior - Brasil (CAPES) - Finance Code 001. National Council for Scientific and Technological Development (CNPq) n° 800673/2016-6 and São Paulo Research Foundation (FAPESP) n° 2014/04748-2, 2015/20525-6 and 2015/12461-8).

**Acknowledgment.** We thank the Center for Semiconductor Components (CCS/UNICAMP) for the use of their equipment.